# C-Ports: A proposal for a comprehensive standardization and implementation plan of digital services offered by the "Port of the Future"

*Paolo Pagano[1], Silvia Antonelli[2], Alexandr Tardo[2]*

*Abstract*—In this paper we address the topic of a possible path to standardize the ICT services expected to be delivered by the so-called "Port of the Future". How the most relevant technologies and Information Systems are used by the Port Communities for their businesses is discussed together with a detailed analysis of the on-going actions carried on by Standard Setting Organizations. Considering the examples given by the C-ITS Platform and the C-Roads programme at EU level, a proposal of contents to be considered in a comprehensive standardization action is given. The innovation services are therefore grouped into four bundles: (i) Vessel & Marine Navigation, (ii) e-Freight & (Intermodal) Logistics, (iii) Passenger Transport, (iv) Environmental sustainability. The standardized version of these applications will be finally labeled as C-Port services. Alongside the standardization plan, a proposal for ranking the ports on the basis of a specially-defined C-Port vector is discussed with the purpose of addressing the well-known lack of consensus around the mathematical definition of the Smart Port Index. Considering the good practice and the background offered by the Port of Livorno in terms of innovation actions, the prospected final user applications are then labeled as Day 1, Day 1.5, and Day 2 services in consideration of the technical and commercial gaps to be filled. As a case study about the evolution in the C-Port vector experienced by the Port of Livorno in the last years will also be discussed.

## I. INTRODUCTION

Seaports are genuine intermodal hubs connecting seaways to inland transport links such as roads and railways. Seaports are located at the focal point of institutional, industrial, and control activities, in a jungle of interconnected information systems.

As seaports operate in freight and passenger businesses, the main vertical applications are in the domain of logistics and digital offer targeted to citizens and tourists. Ports also represent important industrial innovation hubs where local business, together with institutional and public bodies can interact and generate added value services. The valorization, and consequentially the transfer, of scientific and technological results from research centers, hold a crucial and increasingly relevant role in terms of economic development. Indeed, it is considered the driver for the transition from a "manufacturing-based economy" to a "knowledge-based economy".

Traditionally, logistics has been considered as a "non-core process", an activity that does not contribute to the value creation process. In fact it was considered a process based on downward competition and where investments had a tendency to be "limited", if compared to those resources allocated for those main "core" processes.

The concept of "lean", typical approach of the so-called Industry 4.0, based on the streamlining, digitalization, automation and efficiency of the logistics chain upstream and downstream processes can also be applied to other sectors, firstly, to distribution and maritime logistics. Logistics and maritime processes have been redesigned with a view of creating value and eliminating all those wastes imputable to inefficiencies and lack of organization. In this way, ports are becoming increasingly automated and optimized, thanks to the contamination between ICT and robotics as well as to the integration with other attractors located towards the hinterland and oversea.

Although Research and Innovation Actions funded by European Programmes (supporting the development of frontier technologies ranging from 5G to autonomous vessels) do include standardization activities in their implementation plan, an effective and comprehensive attempt to standardize the digital offer by the so-called "Port of the Future" is still missing. Along the same line a standardized metric to rank ports in terms of innovation services is still lacking general consensus.

The authors are therefore considering a bottom-up approach coming from a long-term experience of Research & Development in Livorno, a mid-sized multi-purpose seaport in the Mediterranean Sea.

These activities are framed into the agenda of the Joint Laboratory (JLAB) founded in 2015 under the ownership of CNIT and the Port Network Authority of the Northern Tyrrhenian Sea [1]. The actual target of the two institutions is that of integrating innovative digital services into the offer to the port communities (collective name for ocean carriers, haulers, intermodal carriers, shippers, freight forwarders, insurance companies as well as institutional and control bodies).

To ensure full accessibility and interoperability among the final applications, JLAB is providing a cloud-shaped reference architecture together with a set of standards for data formats and interfaces to services.

In this paper we will show how the strategy followed in the European Union for the large-scale deployment of C-Roads is also applicable to the case of C-Ports.

The description of the piloting experience in Livorno is discussed as an example of a possible starting point for a mid-sized, multi-purpose, innovative port in the Mediterranean Sea.

The authors will present in Section II the reference model borrowed from Intelligent Transportation together with a gap analysis about standardization, in Section III will summarize upon the novelty of this work and anticipate the content included in the following technical sections. In Section IX the authors will drop the conclusions from this work.

## II. REFERENCE MODEL AND GAP ANALYSIS

*A. C-ITS and Smart Roads: a possible reference model*

In the C-ITS domain a comprehensive standardization activity permitted to open the business of Connected and Autonomous Vehicles to the Digital Single Market, in a way that all relevant roads in Europe can follow the same specifications and offer a basic set of services to the most recent high-tech vehicles manufactured by all brands.

The adoption of a standard ICT architecture was originally proposed world-wide through the ISO TC 204 [2] standard committee workplan and then reworked and finalized by ETSI TC ITS [3] and CEN TC 278 [4] implementing the mandate M/453 issued by the European Commission in 2009. The corresponding international standards, ETSI EN 302 665 [5] and ISO IS 21217 [6] have been adopted as the basis for standardizing protocols at all layers, ranging from communications (notably LTE for long-range and IEEE802.11 OCB or cellular V2X for dedicated short-range communications) to data management, security and facility services. In April 2014 ETSI and CEN published a set of 50 documents representing the "Release 1" of the regulations requested by the European Commission [7].

While the Standard Setting Organizations are looking beyond the standards in force tackling with aspects related to autonomous driving (SAE Levels 4 and 5) in the prospected "Release 2" of the standards, e.g., Adaptive Cruise Control, Platooning, Vulnerable Road User Safety, a set of exploitation actions have been rolled out by the European Commission to turn innovation and standardization output into tangible results and scale them up to international large-scale deployment programmes.

EC structured a technical committee, known as C-ITS platform, aimed at developing a "Master Plan for the deployment of Interoperable Cooperative Intelligent Transport Systems in the EU". The Working Groups considered all aspects related to C-ITS ranging from business models to more technical subjects as spectrum allocation, data protection and privacy, and in-vehicle messaging.

The C-ITS platform finally provided a concrete gap as well as a cost/benefit analyses for road safety, traffic information, freight-oriented applications.

The most popular result from the C-ITS platform is the classification of final user services into nine bundles (i.e., safety, motorway, urban, parking, smart routing, freight, vulnerable road users, collision, wrong way), whose services are labeled as "Day 1" and "Day 1.5" services depending on their technical readiness.

The standardization activity together with the stakeholders' recommendations have been endorsed by the European

Commission and turned into funding initiatives. Some Connecting Europe Facility (CEF) [8] calls have been opened to support the C-Roads and the ITS Platform implementation programs at full scale.

Today it is possible to browse the official map of the TEN-T corridors to visualize where the C-ITS equipment is installed all across Europe and what "Day X" services it is providing [9].

*B. Research and development activities in the port and maritime sectors*

The European Commission as well as Member States have funded many innovation projects supporting initiatives oriented both to the development of Key Enabling Technologies (as Mobile Networks and IoT, Digital Platforms and Cloud Computing, Blockchains, Artificial Intelligence, etc.) and to business verticals (as Port Community, Logistics, and Transportation Systems).

Recently port developments have been considered in Horizon 2020, in the "Mobility for Growth" work programme, under the "Port of the Future" topic. The funded projects are entering the ending phase.

The main focus in CEF Programs is about Data Sharing and Interoperability, with an extent of harmonizing formats and easing cross-border logistics.

In [10] the authors review the main technologies and standards being applied in seaport information systems and discuss to what extent and to what verticals GNSS, EDI, RFID, OCR, and wireless communication (including VANET) play a role. The most notable information systems are also discussed therein: National Single Windows, Port Community Systems, Vessel Traffic Services, Terminal Operating Systems, Gate Appointment Systems, Automated Gate Systems, Automated Yard Systems, Port and Road Traffic Information System, Port-hinterland intermodal Information Systems.

How can technologies be framed into standards in order to permit full interoperability among information systems is still debated and the discussion about the integration of the information systems into a unique architecture, capable of offering specialized services for the various verticals, is still ongoing.

*C. Standardization activities in the port and maritime sectors*

While Research and Development is coherently progressing across all aspects, the standardization action is missing a comprehensive work plan.

For what is related to ships and marine technology, without any sake of completeness, we will discuss the scope of notable Standard Setting Organizations such as UNECE, ISO, IMO, and ETSI.

Among the publications by the United Nations Centre for Trade Facilitation and Electronic Business (UN/CEFACT [11]), an intergovernmental body of the United Nations, Electronic Data Interchange for Administration, Commerce and Transport (UN/EDIFACT) comprises a set of internationally agreed standards, directories, and guidelines for the electronic interchange of structured data, between independent computerized information systems. For instance, all events related to container handling and management (as for instance, Gate In/Out in Terminals and Load/Discharge from vessel) are transmitted from haulers and ocean carriers using UN/EDIFACT standards.

Following its mandate, ISO has produced more than 400 standards providing:
- within the work programme of Technical Committee (TC) 8 [12] called "Ships and marine technology", kicked off in 1947: specifications impacting all aspects related to ship construction and operations subject to IMO requirements, for the observation, sailing, and exploration of the sea;
- within the work programme of Technical Committee (TC) 104 [13] called "Freight containers", kicked off in 1961: specifications for freight containers, both general- and specific-purpose, including terminology, classification, dimensions, handling, test methods, and marking (i.e., identification and communication).

IMO is directly maintaining the International Convention for the Safety of Life at Sea (SOLAS) [14], generally regarded as the most important of all international treaties concerning the safety of merchant ships. The document consisting in 14 chapters regulates aspects ranging from fire protection, to life-saving arrangements, and safety of navigation. Radio Communications are considered in relation to maritime distress, requiring passenger and cargo ships on international voyages to rely on satellite communication capabilities (i.e., Emergency Position Indicating Radio Beacons, EPIRBs) as well as on radio equipment such as Search and Rescue Transponders (SARTs).

Apart from participating as stakeholders in IMO standardization activities, ocean carriers have also teamed up in associations like Digital Container Shipping Association (DCSA [15]) to agree on the common technology foundation that enables global collaboration.

Finally, ETSI is actively standardizing:

- within the work programme of Radio Spectrum Matters (RSM) Task Group (TG) Marine [16]: aspects related to radio communication for the safety of life, radio location and navigation equipment; radio access to terrestrial telecommunication networks;
- within the work programme of ETSI ISG on European Common Information Sharing Environment Service and Data Model (CDM) [17]: aspects related to data exchange among different maritime legacy systems of the key strategic objectives of the European Union (transport, environmental protection, fisheries control, border control, general law enforcement, customs and defense) in a cooperative network.

*D. The odd case of Sea Ports*

In the global scenario "sea ports" are considered as pure infrastructures and complexity is resolved in terms of interoperability among application-layer functions. Erroneously, ports are historically considered static realities, with little inclination to change and whose activities are considered accessory and without added value. Information Technology is the main character of the digitalization era, everything is connected in a "smart" way, businesses and their industrial processes have been re-engineered according to the so called "4.0" paradigm.

From an economic assessment, investments in digital port infrastructures have important microeconomic effects, since they result into overall surplus (*welfare*) and, according to long-term growth models, into the economies of scale and specialized production. According to the *New Economic Geography Models* [18], infrastructural upgrading can lead a company located in a specific area to have access to a wider market, as a result of lower transport costs. At the same time, other producers are encouraged to operate in the same area, feeding an increasingly attractive spatial concentration process and increasing the local market competitiveness.

We also know that seaport infrastructures have an important indirect effect on economic growth. That is the reason why national economies consider seaports as crucial business systems operating in a highly competitive market, worth to attract investments in digitalization and infrastructures.

According to the technology setting, "smartness" usually refers to the principles of automatic computing, however, a globally accepted definition of "smart port" does not already exist, as it has not been well defined in the literature, as well as an internationally accepted and standard definition for the word "smart" does not exist in the context of port industries and maritime scenarios [19]. Houston University has defined a Smart Port Index (SPI) to measure ports "smartness". Nevertheless, the different entities may not be easily comparable due to their service and structural heterogeneity, which complicates ports scoring and ranking process [18], [19], [20].

Seaports transfer passengers and goods coming from the ships to other transit points and final destinations complying with international maritime regulations. These regulations affect the procedures and the (digital) administrative transactions allowing a certain person or object to reach the next point of transit.

The main directives in the European Union (mainly) regulate the information set related to the (arrival/departure) schedule time of the ships [21], customs controls for travelers [22], notifications of dangerous freight [23], notification of waste and residuals [24], notification of security to entry ports [25] and entry summary declaration for customs clearance [27]. These regulations sometimes implement more general provisions issued by the IMO to its members and practiced worldwide.

All directives tend to implement electronic exchange of information, avoid duplications, force member states (EU and IMO) to set up a Single Point Of Contact (SPOC) accessed through a National Single Window (NSW) instead of keeping alive dozens of legacy information systems run by independent regulatory and control agencies (see Figure 1).

5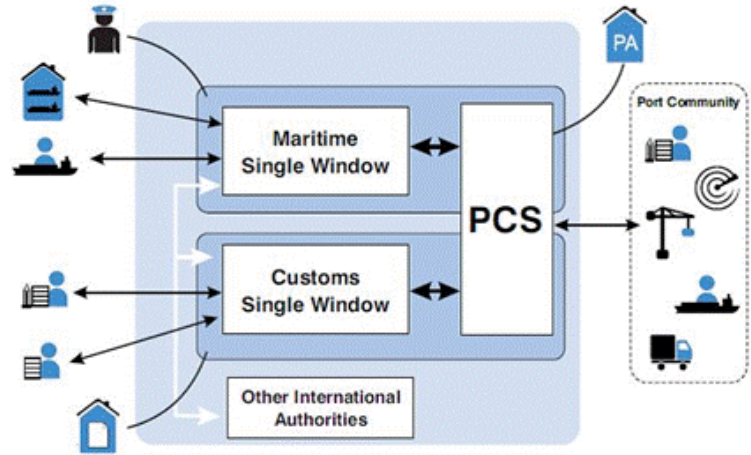

*Figure 1: Concept of national single windows (maritime and customs) handling port operations, customs procedures, technical controls and certification procedures. (Courtesy of Actual I.T. d.d., Koper (SLO))*

Each port is supposed to set up an information system, known as Port Community System (PCS [27]) to connect to the SPOC established at national level; private and public institutions use PCS to enable intelligent and secure exchange of information, to improve the efficiency and competitive role of the seaports, to automate and smooth port logistics processes through a single submission of data and by connecting transport and logistics chains.

PCS are therefore commodity appliances whose scope and technical specifications are delegated to each Port Authority that is responsible of keeping PCS operational towards its users. The Authority is also in charge of upgrading the digital offer to port users extending the scope and integrating new services permitted by KET.

While this approach is opening a genuine competition among European ports pushing them to upgrade their digital service offer, the absence of a comprehensive standardization can be considered as a drawback for all roaming users (i.e., Vessels, Trucks, Trains, and their personnel) that are forced to adapt to multiple port configurations or rely on local specialists for full-customized solutions with a net cost in terms of man-power and redundant software.

Sea ports have strategically joined up in associations and alliances to harmonize Port Community System specifications and guarantee interoperability. To this extent (for instance) the International Port Community System Association (IPCSA) has published an architecture under the name of "Network of Trusted Networks" [28] to share a system for authenticating common users and a set of protocols to share and exchange data (mainly) in import/export processes.

In other approaches, some platforms are assuming the role of "de-facto" standards [29]. In this way data sharing is delegated from local PCS (not standard) to a third-party solution provider (de-facto standard). In the cited solution a permissioned Distributed Ledger Technology platform keeps track of every change in the shared records (i.e., events and documents) so that authenticity, non-repudiation, immutability, and data protection can be guaranteed. As a result, some logistics chains connecting a set of ports, ocean carriers, inland rail-road terminals, shippers, and freight forwarders are already in place especially in Northern Europe.

Not the former nor the latter solutions are opening the port sector to the so-called Digital Single Market as the platform is managed by few stakeholders that are in charge of data custodial and sharing.

*E. The Smart Port Index (SPI)*

Although, "The SPI can facilitate early detection of deficiencies […] to make correctional actions, or help expedite the improvement of port performance. Furthermore, ports can use the SPI to evaluate themselves and know where they stand in comparison to other ports."[19], a unique, globally accepted SPI is not standardized and probably not even possible for the variety of port heterogeneity [19], [20], [30], [31], [32], [33], [35], [36].

Also, many KPIs referenced in literature are subjective and cannot be compared from port to port such easily; it is therefore worth to consider an indicator calculated from C-Port services only.

As the C-Port services are standardized it will be possible to rank all ports on the basis of an indicator having gathered consensus. It will also be possible to profile each port on the basis of its genuine offer without generating artificial outperformances.

A possible classification of C-Port digital verticals can be considered for the following bundles: (i) Vessel & Marine Navigation, (ii) e-Freight & (Intermodal) Logistics, (iii) Passenger Transport, (iv) Environmental Sustainability. Differently from [19] we decided:
- to decouple "Operations" related to logistics from those related to mobility to match the port specificity along one or the other axis;
- to consider "Energy" as one of the aspects related to Environmental Sustainability.

These considerations pave the way towards a comprehensive standardization action whose objectives are illustrated in Sec. III.

## III. CONTRIBUTION OF THIS WORK

In consideration of the missing standardization of port ICT services we hereby:
- propose a service classification into the comprehensive set of the aforementioned vertical domains (see Sec. IV);
- propose a reference model for ranking the seaport in terms of C-Port services (see Sec. V);
- propose to adopt a reference model for the cloud architecture as a convergence stack of all services offered by the "C-Ports" (see Sec. VI);
- propose a definition of service classes as a function of technical readiness and standardization gap as shown in (see Sec. VII).

The technical readiness refers to the availability of (pre-) commercial solutions for networks, digital platforms, and business logic; all of them are indeed the key technologies enabling innovating digital services targeted to the final users.

We will refer to Day 1 services when they are already considered in standardization and can be achieved by state-of-the-art technologies; we will refer to Day 1.5 when they have not yet been proposed in standardization and need to address challenges in technologies; we will refer as Day 2 for services, still beyond state-of-the-art, not yet considered by Standard Setting Organizations and technologically.

On the basis of our experience[3] we consider 5G, IoT, Blockchain (BC), and Artificial Intelligence / Machine Learning (AI/ML) as the most notable Key Enabling Technologies for Logistics and Transportation. We also consider space assets, i.e., Satellite Communications (SatCom), Satellite Earth Observation (SatEO), Satellite Navigation (SatNav) in the recommended set of KETs. In the following we will refer to the above KETs as "reference set".

## IV. SERVICE CLASSIFICATION

From an economic point of view, the port digitization process generates added-value and this has an important redistributive effect for all the stakeholders involved in the port system.

We can say that the entire port community benefits from the port's innovations in two different ways:
1. Directly, in the case of clear improvements in logistics processes in terms of efficiency and optimization, new employment in the port hinterland and GDP;
2. Indirectly, in terms of improved quality of life. For instance, a system of optimization of the yards, based on 5G technologies, can also improve air quality thanks to the reduction of the ships stop time at berth and the port systems optimization (avoiding, first of all, empty trips) and the reduction of accidents.

Research and development activity is, in fact, defined and analysed also in relation to the impact on the end user, who plays an active and proactive role, especially in the definition of requirements.

In the following sections, the scope of the vertical domains is described together with a preliminary list of high-level requirements.

*A. Bundle 1: Vessel and marine navigation*

As already discussed in Sec. II-C marine navigation is mostly disciplined in ISO and IMO regulations.

Ports are not concerned by the Deep-Sea Sailing procedures, as they mostly manage Naval traffic, i.e., the entry (leave) of vessels to (from) port waters. We will therefore focus on procedures and functions related to vessels that are in proximity of ports or docked therein.

The preliminary list of these services will be detailed in the Appendix under the prefix "A", together with the benefit tailored to the port communities. In this context we remark that the bundle will offer (at least) accurate Vessel Positioning (terrestrial and satellite), Low-Rate and High-Rate Vessel-Port bi-directional communication, Full information about cargo, and permit to detect events like "Incident at Sea", "Suspicious Vessel / Maneuver", as well as to provide "Assistance to berth allocation and docking".

---

[3] See for instance the conference programme of the Digital Transport Days 2019 organized by the European Commission in Helsinki in the framework of the public events under the Finnish Presidency of the European Council.

Accurate Vessel Positioning (terrestrial and satellite), Accurate Bathymetric Data, a Real-Time distributed monitoring network (including meteo-marine observations), HD video sources on vessel & port, and a new generation low latency and high throughput terrestrial network are drivers for the release of to this bundle of services.

Some KETs enabling the functions listed above must undertake a full technological transfer process. Notable examples from our "reference set" are:
- 5G for High-Rate/Real-Time Vessel-Port bi-directional communication;
- IoT for Accurate Bathymetric Data, Real-Time meteo-marine monitoring, HD video sources on vessel & port;
- BC for retrieving reliable information about cargo;
- AI/ML for data aggregation, and on-line analytical processing.

*B. Bundle 2: e-Freight & (Intermodal) logistics*

Container dimensions, classification, and sealing are fully standardized. Communications related to the management of containers in port terminals have been standardized since the '70s by the aforementioned UN/EDIFACT. On the contrary, the specifications related to general cargo is loosely standardized.

Other aspects, from data aggregation to event building, securing and sharing, is left to the open market so that stakeholders can rely on efficient platforms and services for tracking and tracing of goods along the (intermodal) logistics chain.

Ports are the pivot place to connect Line Operators to the terminals (this is the original scope of the Port Community Systems), as well as to the interconnect port gates with the smart roads, specifically those included in the TEN-T network.

The preliminary list of these services will be detailed in the Appendix under the prefix "B", together with the benefit tailored to the port communities. In this context we remark the bundle will offer (at least) Freight Management and Control, Gate Automation, In-port Smart Navigation.

Distributed monitoring network, data aggregation, and on-line analytical methods applied to containerized and general cargo pervasive monitoring and control in port areas (docks, warehouses, stores), Automatic identification of users, vehicles and goods, Real-time communication between port and logistics means, Port-to-Port / Port-to-Road / Port-to-Railways communications, are drivers for the release of this bundle of services.

Notable examples of technologies (from our "reference set") undertaking a technological transfer process for the functions above are:
- 5G for implementing real-time communication Port-Terminals-Trucks;
- IoT for implementing a distributed monitoring network, pervasive monitoring and control of freight in port areas (docks, warehouses, stores), automatic identification of users, vehicles and goods;
- BC for retrieving reliable from port to other inland attractors;
- AI/ML for data aggregation, and on-line analytical processing.

*C. Bundle 3: passenger transport*

Ports are genuine intermodal Points of Interest. The future of the maritime mobility passes through digital innovation in ports. Ports need to manage up-to-date timetables collecting information coming from ferry and cruise lines as well as from VTS. To ease in-land mobility, ports are expected to interoperate with Train and Road Traffic Control Centers (TCC). Allowing such kind of functions, passenger transport in seaports will contribute to the implementation of the Mobility as a Service (MaaS) concept, considering the integration of various forms of transport services into a single mobility service accessible on demand [37]. MaaS is a human-centered approach aiming at providing to citizens access to multimodal mobility services, single journey planning and ticketing options for the user, as well as the provision of reliable and advanced travel information from the planning phase until the end of journey. The MaaS concept will be boosted by Cooperative, Connected and Automated Mobility (CCAM) technologies as the vehicles will be more and more considered as "means" (i.e., to be rented, handed to/from somebody else) rather than a personal commodity. In the C-Ports passengers will be dropped in front of the booked ferry by automated vehicles.

The preliminary list of these services will be detailed in the Appendix under the prefix "C", together with the benefit tailored to the port communities. In this context we remark the bundle will offer (at least) Infomobility and journey monitor, Integration with TCC of the C-Roads and Railways, and In-port Smart and Autonomous Mobility.

Journey planner and manager (for booking and payment), Just-In-Time information delivery, Port-to-road full-fledged data exchange, Port-Vehicles-Pedestrians real-time communication, are drivers for the release of to this bundle of services.

The most relevant gap to deliver C-Port services in this bundle refers to the adaptation of existing protocols and possible integration of independent information systems (as TCC and C-ITS). Services related to Port-Vehicles-Pedestrians real-time communications require a pervasive 5G network, still within the scope of state-of-the-art technologies but not yet realized in full scale.

*D. Bundle 4: environmental sustainability*

Starting from the Kyoto protocol of 2005 and moving towards the recommendations coming from the Paris Climate Agreement of 2015, seaports are expected to adopt regulations and policies aimed at mitigating their carbon footprint.

We have demonstrated how technology development is a key element for the sustainable growth of the port [38].

It is envisioned to have a dashboard capable of ranking docks, berths, terminals, and areas, in terms of pollution levels and more specifically carbon footprint.

The preliminary list of these services will be detailed in the Appendix under the prefix "D", together with the benefit tailored to the port communities. In this context, we remark the bundle will offer (at least) Pollution Control (including COx and noise), Road Traffic Level Control, and Dynamic pricing (of all services) to Line and Terminal Operators.

A full-fledged distributed monitoring network, data aggregation, on-line analytical processing, data mining and knowledge extraction capabilities, are drivers for the release of to this bundle of services.

Notable examples of technologies (from our "reference set") undertaking a technological transfer process for the functions above are:
- IoT for implementing a distributed monitoring network;
- BC for storing and securing certified data series;
- AI/ML for data aggregation, on-line analytical processing, data mining.

## V. THE C-PORT VECTOR

*A. Mathematical formulation*

As a unique comprehensive metric is welcome to rank ports for their ICT innovation potential, we propose the following C-Port Vector whose i-th component is defined as:

$$\text{C-PV}_i = \rho |a_i C_{i,j}| \times |w_j| = \rho \cdot \begin{vmatrix} a_{Nv}P_{Nv} & a_{Nv}D_{Nv} & a_{Nv}R_{Nv} \\ a_{Fr}P_{Fr} & a_{Fr}P_{Fr} & a_{Fr}R_{Fr} \\ a_{Mb}P_{Mb} & a_{Fr}P_{Mb} & a_{Fr}R_{Mb} \\ a_{St}P_{St} & a_{St}P_{St} & a_{Fr}R_{St} \end{vmatrix} \cdot \begin{pmatrix} w_P \\ w_D \\ w_R \end{pmatrix}$$

$$(i = Nv, Fr, Mb, St)$$

*Equation 1*

where:

- ρ is defined as the standardization merit factor, calculated as the ratio of adopted standards on the applicable standards;
- $C_{i,j}$ is the port innovation matrix where the element represents the total cumulative cost in unit of M€ of project work expenses for developing new Prototypes (TRL ≤ 5), new Demos (6 ≤ TRL < 8), and new Released applications (TRL ≥ 8) in the four C-Port application areas (Navigation – Nv, Freight – Fr, Mobility – Mb, Sustainability – St);
- the $|a_i|$ vector represents the port specific businesses (either freight, or passengers, or energy) where the navigability is considered higher for stormy or narrow ports than in accessible ones. A normalization constraint applies as $\sum \frac{1}{a_i^2} = 1$;
- the $|w_i|$ vector represents the innovation reward, i.e., the weight assigned to early development of innovative technologies in respect to mature deployments. A normalization constraint applies as $\sum \frac{1}{w_i^2} = 1$;

If detailed records about innovation projects are not available it is always possible to quote the total investment summing all elements in the port innovation matrix:

$$Tr(^T\tilde{C} \cdot \tilde{C}) = \sum_i \sum_j a_{ij}^2$$

*Equation 2*

where $\tilde{C} = \left[\sqrt{a_{ij}}\right]$ is introduced to get rid of the quadratic summation.

### B. Properties of the C-Port vector

The metric defined in Equation 1 shows the following properties that render it robust and eligible for global usage:

1. the C-Port Vector is null if no standard is adopted; two ports with similar innovation potential can result into C-Port Vectors with very different modules if one of them relies on proprietary solutions;
2. projects are weighted in terms of their cost as it is hard to consider other performance indicators (like publications, number of new tenure positions, etc.) without generating any systematics;
3. different ports have different C-Port Vectors in the 4-dimension space so that it is possible to define the scalar product of C-Port Vectors representing the port vocation (as freight, passengers, energy, etc.), as well as to compare the effort devoted to innovation along the 4-axes. The generic angle between two C-Port vectors is defined as:

$$\alpha = \cos^{-1}\left(\frac{C\text{-}PV_1 * C\text{-}PV_2}{|C\text{-}PV_1| \cdot |C\text{-}PV_2|}\right)$$

*Equation 3*

Of course, the two vectors can represent two snapshots of the same port taken in different timing. This will be investigated as Case Study in Sec. VIII.

4. the innovation reward is a tunable parameter that can privilege ports connected to innovation realms (like research campuses and universities, start-up's, etc.) featuring early-stage digital services or those having already released new services to the user communities.

## VI. THE INNOVATION PROCESS AT THE PORT OF LIVORNO

The Livorno port, favorably situated in western Tuscany, plays a major role in the European internal trade, ensuring frequent and fast maritime connections to southern European countries, as well as in the EU external trade, thanks to its well-established linkages to northern African countries and the Americas. Livorno is also considered a pivotal node in the logistic chains linking the Mediterranean to central-east Europe. The Port Authority in collaboration with JLAB designs and supervises the implementation of the Monitoring and Control Architecture (MoniCA), the Port Information System [39].

For the sake of this implementation the activities are oriented to standardization and interoperability, as well as to the upgrade of port ICT assets related to Sensing, Networking, Data Management, and Cyber-security.

Within MoniCA all the processes orchestrated by the Port Authority take place, including all what is related to passenger mobility (e.g., MaaS) and freight logistics (e.g., gate transit and e-custom). These strands converge into a very challenging Digital Agenda funded in part by the EU R&D programmes.

The Authority has therefore invested into the standardization of the reference cloud stack to be considered in current and future tenders and procurements; all innovation processes in the work plan of the collaborative projects are integrated in the (experimental) offer of digital services targeted to citizens and industrial communities.

### A. The MoniCA reference stack

The architecture represented in Figure 2 is structured as a private cloud with a full decoupling of the three canonical layers [40].

While fulfilling a high degree of scalability/flexibility, the system permits to keep compatibility with existing Port IT systems, most notably the Port Community System and the vessel monitoring services (port interface to the National Maritime Single Window managed by the Coast Guard).

The implemented architecture allows the development and release of applications in the shape of microservices, exposing them through special interfaces (APIs) in order to make them usable.

Indeed, microservices represent an architectural approach for the creation of applications. What distinguishes the architecture based on microservices from traditional monolithic approaches is the software description into functional and autonomous components. Each function (effectively a microservice) can be compiled and implemented independently. Therefore, individual microservices can work (and eventually fail) without compromising the others.

This architectural approach permits the development of new services by independent companies avoiding conflicts on IPR and reducing the needs to access a general common background.

All this development is subject to continuous integration and assessed against best-in-class cyber-security requirements [41][42][43]. At the Infrastructure layer the architecture features full support for IoT devices considered attached either to low-rate personal and local area networks (e.g., connected through IEEE802.11 and IEEE802.15 families of standards) or to new generation mobile networks (i.e., 4G and 5G) as producers of Massive Machine-type Communications.

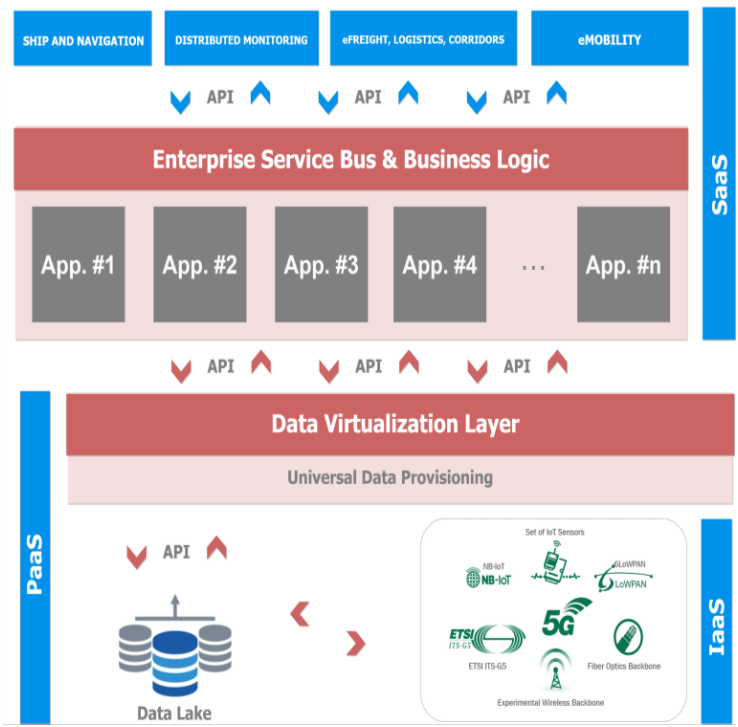

*Figure 2: The MoniCA functional stack.*

At the Platform layer the intention is to establish a Data Lake for controlled and permissioned access appropriately profiled to stakeholders and personal users. In the Data Lake accurate and trusted information about (multi-modal) transport & logistics, environmental conditions, status of operational processes can be retrieved (see Figure 3).

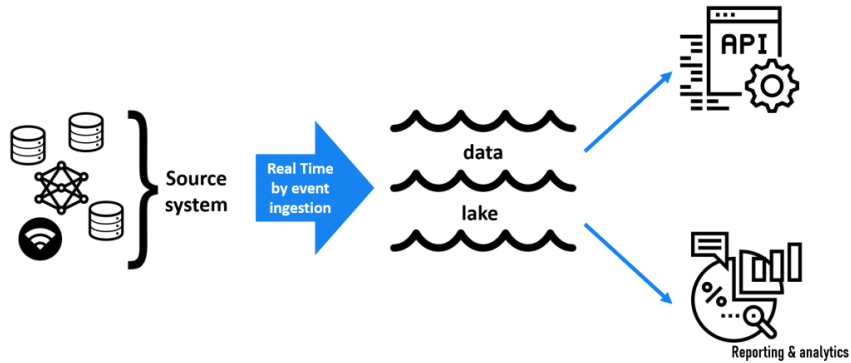

*Figure 3: The data lake concept in MoniCA.*

As an implementation of the Data Lake, a Data Virtualization functional block offers a Universal Data Provisioning service towards data sources. This block: (i) enforces the rules of authorization around data access and retain records about data origin; (ii) capture syntax/semantics of data source schemas and dynamically observe changes over time; (iii) hide implementation differences from the underlying technology (either Relational DBMS, or M2M, or GIS, or document-based repository); (iv) provide caching capability combined with low latency real-time query optimization.

Special attention has been devoted to IoT sources. IoT sensors are kept generic with the requirement to comply with a data model maintained by JLAB and evolving following the specifications published by ETSI TC SmartM2M and OneM2M alliance. The data custodial of IoT data is therefore implemented by a M2M compliant platform storing the data streams coming from distributed sensors in indexed containers.

Still at Platform layer the ICT stack is interoperated with a set of external Distributed Ledgers to be able to collect and share information in a secure and non-repudiable way. The solution, called "Interoperachain", decouples from actual DLT thus avoiding technology lock-in from specific proprietary solutions.

At Software layer, a commercial implementation of the Enterprise Service Bus (ESB) concept allows to decouple the application level from the back-end one. The needed functionalities are related to:

- API Gateway: allowing to create APIs (in the form of a list of services), publish them (managing the entire life cycle), connect them to a back-end having collected all the parameters involved in the related calls;
- Single-Sign-On for user management. It allows users to manage the invocation of the related APIs;
- a high-performance queue, enabling communication between different microservices as applications may consist of different microservices that need direct communication in order to deliver the required functionality.

Our actual instance of the MoniCA stack consists of:

- a set of VM managed by VMware;
- a Data Lake implemented by Jboss Teiid, virtualizing actual instances of mySQL, MS SQL, Ocean Mobius, MongoDB, ESRI ArcGIS;
- WSO2 as open-source suite for the development of micro-services and integration of third-party applications;
- Docker is used to implement microservices with container-design pattern.

The innovation services described in Sec. VI are integrated into this stack and hosted in the JLAB computing farm described below. As a notable example of service running on the MoniCA stack, Livorno relies on a fully operational Port Community System with more than 600 users [44].

### B. The staging and validation process

A staging farm is installed in the JLAB premises. Every innovative ICT service is considered as unit of code and must follow an integration testing process. Indeed, after the development phase by third-party companies, the new service is hosted in a secure industrial tenant where conformance and interoperability tests are performed.

In Figure 4 the components of the staging and testing environment are shown. A capillary network encompassing the fiber-optics backbone and the wireless networks (WLAN and 4G/5G mobile networks) is connecting the topmost cloud layers with distributed sensors and actuators in port waters and on the landside. The staging and validation environment has been extensively used in research and demonstration activities as well as for validating new services procured by the Port Authority and implemented by third parties. Some examples will be shown in the next section.

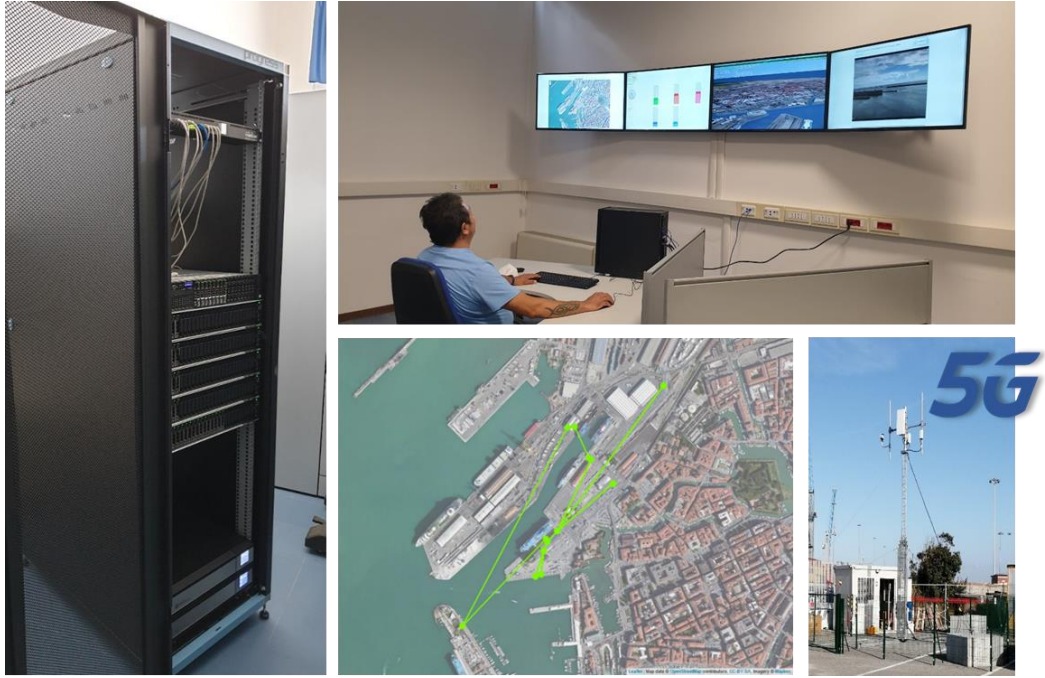

*Figure 4: The ICT infrastructure, testing environment and command center in Livorno.*

## C. Other relevant innovation examples in Europe

Although not actively contributing to a comprehensive standardization plan (as discussed in Sec. II.D), the major seaports are tackling with the topic of digital innovation offering areas for testing and experimentation, hosting laboratories, organizing hackathons, incubating start-ups in the so-called innovation districts.

Without any sake of completeness, we report hereby some remarkable actions taken by Port Authorities to foster innovation in port businesses:

- Antwerp who has invested in "Nxt Port" [45] for data sharing among communities, has prototyped a digital twin of the port [46], has started an innovation district active in shipping and logistics matters [47];
- Rotterdam is active in gathering innovation proposals through the "Smartest Port" initiative [48];
- Hamburg has started an initiative similar to the former called "Homeport" [49].

The approach is always that of using the port as a large-scale testbed in order to assess the applicability and the readiness of certain innovative services that the authors have classified as in Sec. IV. To this extent the experiences reported in Sec. VII cannot be seen as unique in EU but represent a comprehensive assessment of experimental results in the full spectrum of the C-Port standards.

## VII. C-Port prototype services in Livorno

Application development is out of the scope of JAB activities. The major part of available innovation services has been developed as proof of concept in EU funded actions. They can be qualified from TRL 6 ("Field Demo of Subsystem Demo Prototype") to TRL 7 ("System Prototype Demo in Operational Environment") depending on the level of integration with the ICT reference stack described in Sec. VI.A.

### A. Vessel and marine navigation in Livorno

In the context of "Bundle 1" (see Sec. IV-A) Port of Livorno has gathered experience in seabed monitoring and modeling [50] as well as in radar-based vessel detection in port waters [51] (See Figure 5**Errore. L'origine riferimento non è stata trovata.** and Figure 6).

The achievements have paved the way to:

- the continuous update in the cartographic support (annotated with up-to-date bathymetric data) provided to active pilots, vessel crew, coast guard, and port authority personnel. A first hint to realize a "digital twin" of the seaport where the vessel is entering (leaving). This prototyped service is enabled by M2M technology (i.e., the bathymetric probe) for data custodial at platform layer.

- a photonics-based radar network detecting passive boats in a more efficient way than traditional radar; the new radar network, integrated with surveillance images from dedicated cameras enables an early and effective identification of the unauthorized approach to the port entrance. This prototyped service is enabled by new IoT devices (i.e., the photonic radars and the cameras) and M2M technology for data custodial at platform layer.

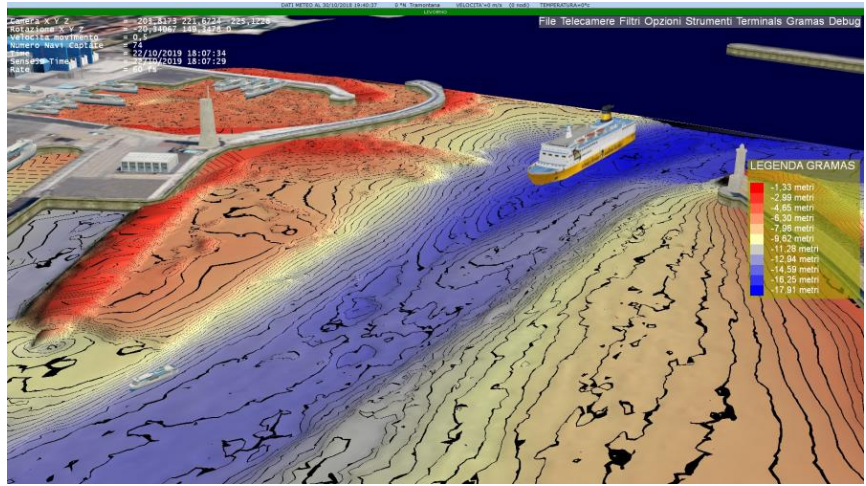

*Figure 5: An instance of the 3d live map of the Livorno seaport with annotated bathymetry.*

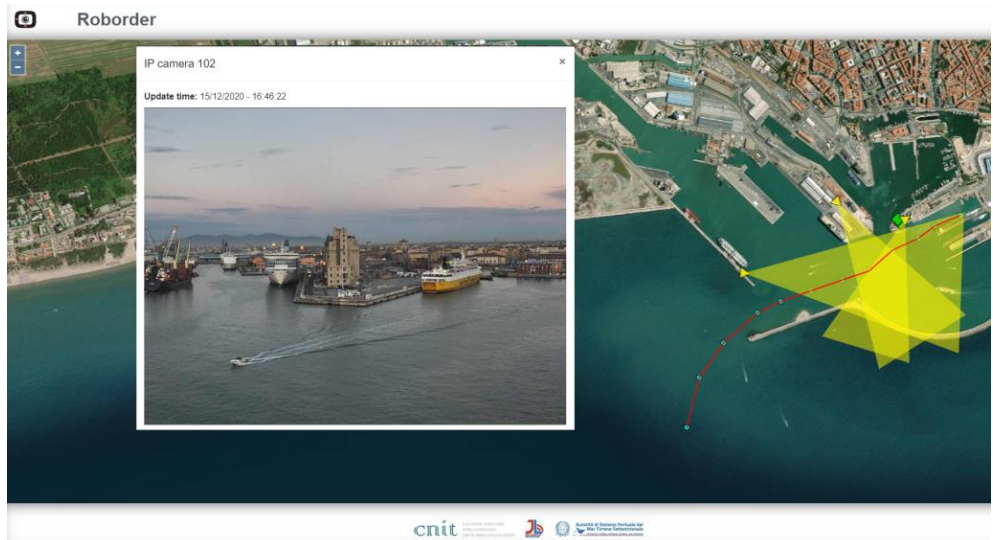

*Figure 6: An example of combined tracking with photonic radar aggregated with AIS track and HR web cameras.*

Following this experience, we propose the classification of A-services as reported in the Annex. Two innovative services are considered ready to be deployed (A.1 – Vessel Traffic Management and A.5 – Berth allocation and docking) thus labeled as Day 1, two others (A.3 – Incident at Sea and A.4 – Suspicious Vessel / Maneuver) are labeled as Day 1.5, as they are still awaiting AI/ML and full 5G coverage. A.2 – Vessel maneuvering is still considered beyond the state of the art and labeled as Day 2.

*B. e-Freight & (Intermodal) logistics in Livorno*

In the context of "Bundle 2" (see Sec. IV-B) Port of Livorno has gathered experience in Cooperative ITS [52] as well as in 5G-based freight management [53] (See Figure 7 and Figure 8).

The achievements have paved the way to:

- deploy a C-ITS infrastructure enabling the services of bottleneck removal (real-time information and early notification about potential traffic congestion, accompanied by suggestion of alternative routes), safety information (real-time information about hazard detected ahead on the road) and smart truck parking (truck drivers are suggested to use smart parking premises for a time lapse optimized on the basis of the real-time traffic along the route and the operational status at the port of Livorno).

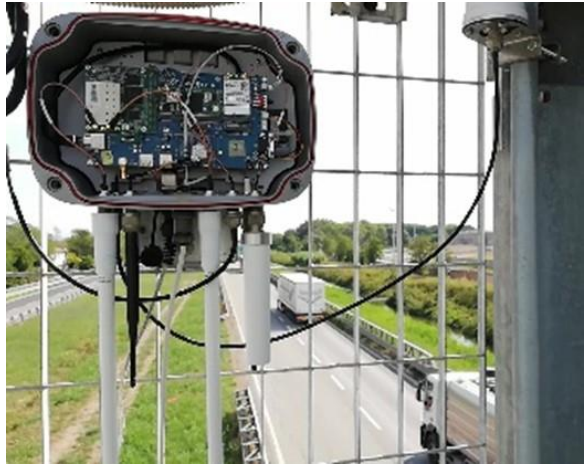

*Figure 7: A Road Side Unit supporting C-Road Day 1 services installed along the Florence – Livorno freeway.*

- develop a 5G-based Model-Driven Real Time Module, allowing a better management of the general cargo (e.g., storage optimization, yard-vehicles call optimization, loading/unloading phases optimization, etc.), resulting in faster throughput compared to traditional human-driven communications.

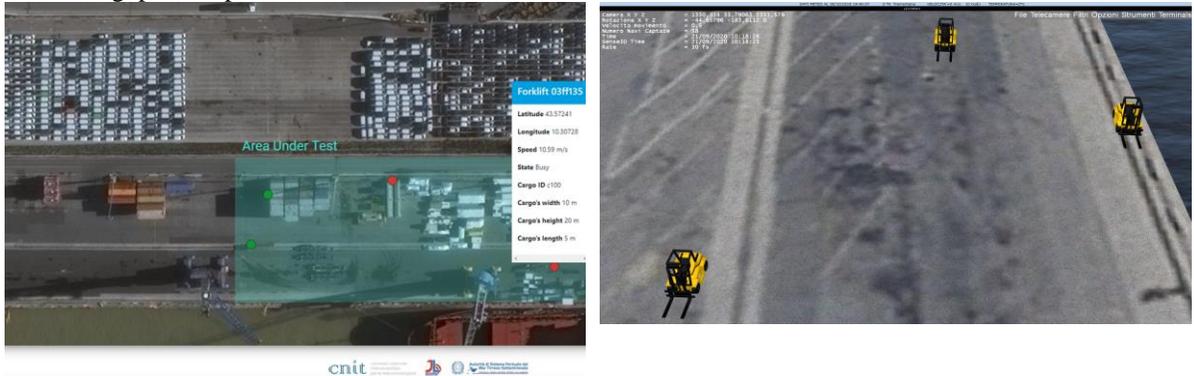

*Figure 8: The yard management system prototype connecting the RTPORT functions (implemented by Ericsson) and the Port Information Stack. In the figure the job assigned to a forklift in dock used for break-bulk freight. The forklifts position in real time in the docks (rendered in 3d VR).*

Following this experience, we propose the classification of B-services as reported in the Annex. Two innovative services are considered ready to be deployed (B.1 – Freight Management and Control and B.3 – In-port Smart Navigation) thus labeled as Day 1, while the others (B.2 – Gate Automation, B.4 – Freight Routing, and B.5 – Incident at Landside) are labeled as Day 1.5, as they are still awaiting AI/ML and integration activity between different vertical functions.

15*C. Passenger (Intermodal) transport in Livorno*

In the context of "Bundle 3" (see Sec. IV-C) Port of Livorno has gathered experience in Autonomous Driving [54] as well as in passenger maritime mobility [55].

The achievements have paved the way to:

- deploy an IoT infrastructure capable of providing information to the vehicle from the edge (C-ITS) and the platform layers (M2M platform) of the Port Information System. Autonomous driving functions (e.g., Vehicle Breaking and/or Lane Change) along the highway (e.g., detected obstacle in front) have been complemented with urban driving Use Cases as those related to the safety of vulnerable users of the port areas (i.e., cyclists and pedestrians).

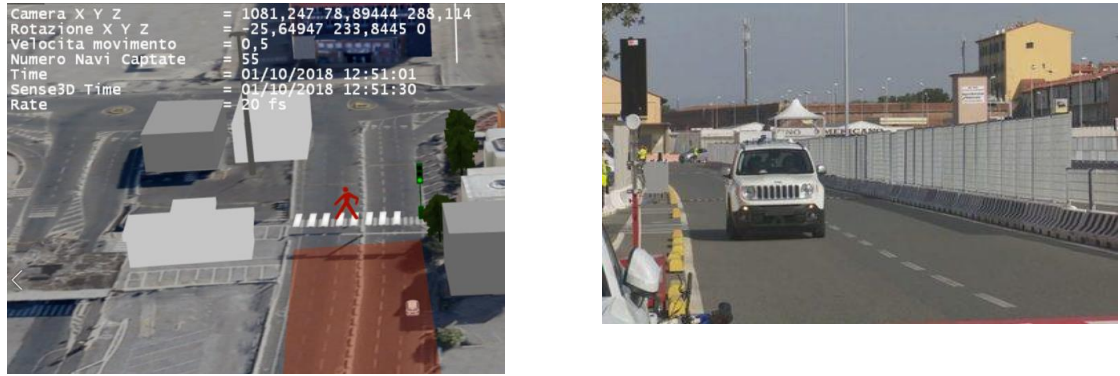

*Figure 9: An event of red-light violation by a pedestrian (rendered in 3d VR) with an autonomous vehicle stopping in front of the green light (from AUTOPILOT demo session in Livorno, on Oct. 19th 2018).*

- develop a digital platform which will collect and integrate all the public transport dynamic and static data (both scheduled and real-time ones) with a proper register tracking all the events. Data are formatted in standard formats (i.e., GTFS, GTFS-RT, EU-Transmodel-NeTEx) and will be broadly available without restrictions, according to the "open data" paradigm.

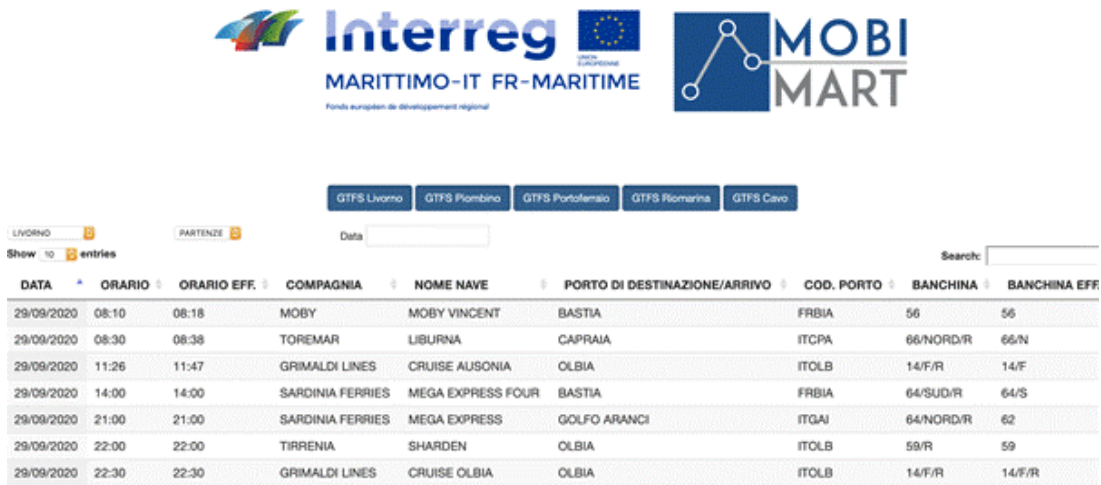

*Figure 10: The timetable for Port of Livorno (and minor ports in the same organization) with interfaces to open-data in various formats.*

Following this experience, we propose the classification (reported in the Annex). All innovative services (C.1 – Infomobility and journey monitor, C.2 – Integration with TCC, C.3 – In-port Smart and Autonomous Mobility (including safety)) are labeled as Day 1.5. The gap is mostly related to missing standardization in the interoperability between TCC and the C-ITS as the existing service prototypes are not going beyond the pre-commercial stage (TRL 5).

16*D. Environmental sustainability in Livorno*

In the context of "Bundle 4" (see Sec. IV-D) Port of Livorno has gathered experience about how can 5G enable good practices to undertake a trend of sustainable growth [53] (Figure 11).

The achievements have paved the way to:
- deploy an environmental dashboard capable of calculating the formation of atmospheric pollutants (as well as noise levels) emitted by the ships during the arrival, internal movement, stop and departure phases at the port of Livorno. The dashboard distinguishes the contribution of individual docks and berths.

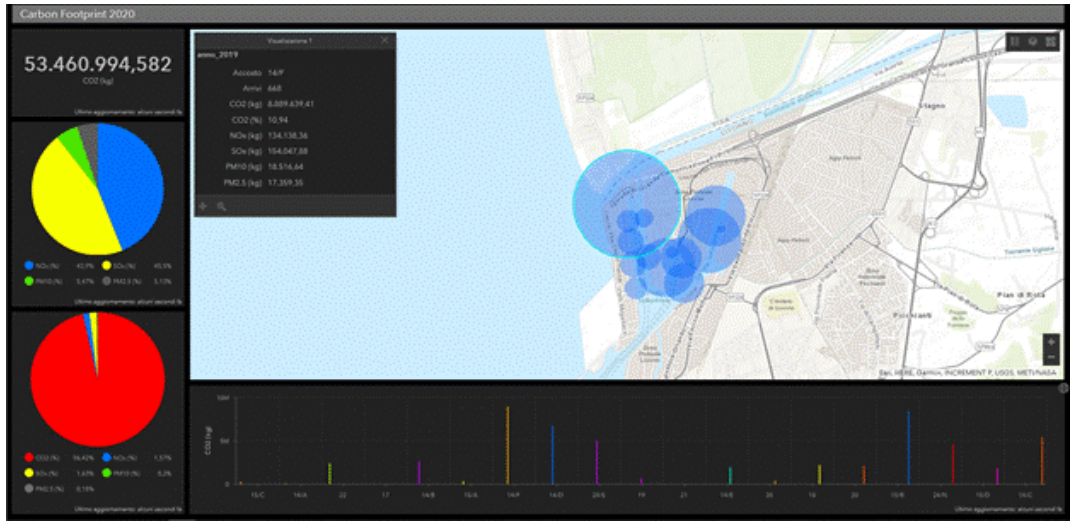

*Figure 11: A snapshot of the environmental dashboard implemented for the Port of Livorno.*

- develop a model to calculate the reduction of CO2 emissions coming from the optimization of yard management operations. Just in consideration of the reduced operating hours of tower cranes and forklifts, Port of Livorno can reach 8.2% of CO2 saving [38]. More savings will come from optimizing vessel's berthing time and freight transfer towards other inland nodes in the logistics chain.

Following this experience, we propose the classification (reported in the Annex). Two innovative services (D.1 – Pollution Level (including COx and noise) and D.2 – Road Traffic Level) are labeled as Day 1.5, where the gap is mostly related to missing capabilities in data aggregation and knowledge extraction. D.3 – Dynamic pricing (all services) to Vessels, Terminals is still considered beyond the state of the art and labeled as Day 2. In this case much more than ICT technology, the service needs a new revolutionary (and green) approach in the port policy for selecting haulers and ocean carriers.

VIII. DYNAMICS OF THE C-PORT VECTOR IN LIVORNO

The C-Port vectors for the port of Livorno have been calculated by considering the Port Authority investments in innovation (reported in Table 1) during two biennia [56]. Respectively C-PV$_1$ refers to 2017-2018 and C-PV$_2$ refers to 2019-2020.

*Table 1: Livorno Port Authority investments in innovation (2017-2020), by areas of interest (bundles). The budget allocated to each action is fully accounted in the first year.*

|  | **Vessel and marine navigation** | **e-Freight & intermodal logistics** | **Passenger transport** | **Environmental sustainability** | **TOTAL AMOUNT** |
|---|---|---|---|---|---|
| **2017** | 137 k€ | 773 k€ | 82 k€ | 310 k€ | **1304 k€** |
| **2018** | 269 k€ | 435 k€ | 770 k€ | 3,447 k€ | **4922 k€** |
| **2019** | 444 k€ | - | - | 492 k€ | **937 k€** |
| **2020** | - | 71 k€ | - | 80 k€ | **151 k€** |

As the same attention has been devoted to require the compliance to the standards where available we consider ρ (as defined in Sec. V.A) constant in the two biennia. The α angle between the two C-Port Vectors, calculated from the definition of the scalar product in $\mathbb{R}^4$ (see Equation 3), is equal to 35°.

This is not surprising as the innovation agenda has deeply changed from 2017 to 2020 [57] where more applications have been submitted in the vertical sector of vessel navigation. The $\alpha$ angle would have been null in case of continuity, or very different from zero in case of sensible changes in the innovation agenda (Figure 12).

Pictorially we render the dynamics of the C-Port Vector as it follows:

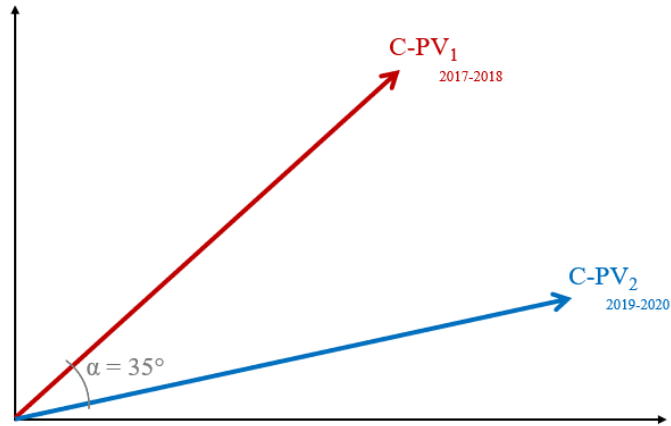

*Figure 12: Dynamics of the innovation at the Port of Livorno estimated from the evolution in time of the C-Port Vector.*

*Table 2: Individual bundles as a percentage of total new investments*

|           | **Vessel and marine navigation** | **e-Freight & intermodal logistics** | **Passenger transport** | **Environmental sustainability** |
|-----------|----------------------------------|--------------------------------------|-------------------------|----------------------------------|
| 2017-2018 | 0,01                             | 0,09                                 | 0,04                    | 0,86                             |
| 2019-2020 | 0,37                             | 0,01                                 | -                       | 0,62                             |

Table 2 shows the weight of each bundle (component) on the total investment. It is noted that Livorno is an ancient port (founded in the XVI century) with many navigability limitations in adverse meteo-marine conditions. This is pushing the administration to invest in navigation and maneuvering machine-aided systems to improve the 24/7 accessibility of the maritime node.

## IX. CONCLUSIONS

In this paper we have demonstrated how an actual comprehensive standardization plan for C-Ports at the European scale is fully achievable starting from the most innovative experiences in the field. Indeed, new digitalization processes can be classified as ready to be implemented (Day 1) whereas the large majority is missing a last mile integration (Day 1.5). Only a few others can be considered beyond the state of the art (Day 2). Ready and not-ready services should rely on standardized specifications to foster market competitiveness among seaports and their digital service providers.

As a side-result of the comprehensive standardization process, a mathematical framework to rank ports on the basis of the ICT innovation processes can be formulated. If applied to the case study of Livorno (presented through a gallery of technical results in the scope of the four appointed verticals), we demonstrate that our metrics represent the dynamics of the technical agenda implemented by the Port Authority [57]. In the case study we have shown how the interest is switched from traditional investment to new challenging technologies not fully exploited so far.

This result can pave the way to extensive research on digital ports with the achievement of offering a common basis for new innovation services (i.e., the C-Port services) that can be quoted to rank one port in respect to the others, irrespectively to the port heterogeneity that so far has rendered impossible such a comparative assessment.

If adopted at the EU scope, this model can also feed a rolling plan similar to that of the C-Roads [9]. The players will be capable in a not-so-far future to browse the Map of Europe with a rendering of C-Port services, offered by each maritime node from the standardized set.

## DECLARATION OF COMPETING INTEREST

The authors declare that they have no known competing financial interests or personal relationships that could have appeared to influence the work reported in this paper.

## ACKNOWLEDGMENT

This work presents results funded in part by the European Commission through the collaborative projects [50][51][52][53][54][55]. The author would like to thank the Port Network Authority of the Northern Tyrrhenian Sea for the appreciation of CNIT work in the last five years as well as all JLAB members who contributed to the results discussed above. The author would also like to thank Ms Lina Konstantinopoulou (EURORAP) and Mr Sergio Escriba (INEA) for the interesting conversations motivating this paper.

19APPENDIX

A preliminary classification of services on the basis of the experience in Livorno is discussed in Sec. VI and reported in Table 3.

*Table 3: Preliminary classification of C-Port Services in terms of technological readiness.*

| CODE | ENABLING FUNCTIONS | MISSING FUNCTIONS (Notes) | STAKEHOLDERS AND PARTNERS (Port community) | BENEFITS |
|---|---|---|---|---|
| DAY 1 | | | | |
| A.1 Vessel Traffic Management | Accurate Vessel Positioning (terrestrial and satellite), Full information about cargo, Low-Rate Vessel-Port bi-directional communication | None | SHIPPING COMPANIES | Real time tracking and monitoring, both at sea and at berth can be useful to calculate an accurate ETA/ETD and to prevent time losses and unpleasant situations (such as the Ever Giver accident in the Suez Canal, which caused huge delays and profit losses in the whole global freight traffic) |
| | | | TERMINALS | Real time vessel positioning and cargo details are useful to define accurate terminal operations |
| | | | FREIGHT FORWARDERS | Real time full information about cargo, both at sea and in the port (e.g. full information about terminal operations and handling) is important for the logistics flow management. It is know that delays demurrage cause huge extra-costs, thus, deadlocks prevention is severely recommended |
| | | | INSURANCE COMPANIES | In case of cargo damage, fires or accidents at sea, accurate vessel positioning, full information about cargo and Vessel-Port bi-directional communication (share of digital resources) is important to reconstruct what happened. Then, the insurance |
| | | | | company will compute the right insurance claims |
| | | | PORT AUTHORITY | Real time and reliable cargo ship information and accurate vessel positioning is full available |
| | | | COAST GUARD | Vessel traffic management and control |
| A.5 Berth allocation and docking | Accurate Vessel Positioning (terrestrial and satellite), Accurate Bathymetric Data, Low-Rate Vessel-Port bi-directional communication | None | SHIPPING COMPANIES | Real time tracking and monitoring, both at sea and at berth can be useful to calculate an accurate ETA/ETD and to prevent time losses and unpleasant situations |
| | | | TERMINALS | Real time vessel positioning and cargo details are useful to define accurate terminal operations |
| | | | INSURANCE COMPANIES | In case of accidents and damages happened during the berthing and docking operations, a full tracked berthing operation is very important to establish contractual responsibilities in the event of a claim, such as ship hull damages or groundings |
| B.1<br><br>Freight Management and Control | (Containerized and General) cargo pervasive monitoring and control in port areas (docks, warehouses, stores). | None | LOCAL SMEs | Real time cargo monitoring is important for the post-sales customer service: producers can easily track their products both at sea and on the other steps of their specific supply chains |
| | | | FREIGHT FORWARDERS | Real time full information about cargo, both at sea and in the port (e.g. full information about terminal operations and handling) is important for the logistics flow management. It is know that delays demurrage and de cause huge extra-costs, thus, deadlocks prevention is severely recommended |
| | | | HAULIERS | Real time information is useful to optimize the port area access time slots |
| | | | TERMINALS | Accident prevention and optimized forklift allocation, avoiding empty trips, and cargo pervasive monitoring in port areas, including the stocked cargo full monitoring |
| | | | INSURANCE COMPANIES | Freight management and control operations are useful to understand, in case of fire, damage or cargo perishment, the amount of the insurance compensation and who are the responsible actors according to the real causes and the real accident dynamics |
| | | | CUSTOM OFFICES | Thanks to blockchain technologies, reliable and trusted information about cargo is available and helps the Custom Office operations organizing freight border control, avoiding time wasting due to document checking. |
| B.3<br>In-port Smart Navigation | Real-time communication Port-Terminals-Trucks | None | SHIPPING COMPANIES | In-port smart navigation helps hauliers to optimize their trips and to coordinate their activities according to terminal and ship availability, avoiding dead times, according to safety and security requirements |
| | | | HAULIERS | Real time information is useful to optimize the port area access time slots |
| | | | INSURANCE COMPANIES | In-port smart navigation systems avoid accidents mitigating the risk of safety and security claims, defining a safe and secure path |
| **DAY 1.5** | | | | |
| A.3<br>Incident at Sea | Accurate Vessel Positioning (terrestrial and satellite), IoT-based distributed network | Data aggregation and on-line analytical processing | SHIPPING COMPANIES | An accidents at sea prevention system is strictly recommended to prevent monetary losses due to damages and delays |
| | | | FREIGHT FORWARDERS | Cargo perishment prevention, using sensorized containers and other IoT technologies, allow an immediate repairing action in order to prevent cargo irreversible damages. According with this, a reefer cargo monitoring platform will be supplied |
| | | | INSURANCE COMPANIES | Adopting smart systems to prevent vessels and-or cargo accidents at sea, in order to calculate an adequate insurance premium and to prevent extra-costs deriving from insurance indemnities |
| | | | COAST GUARD | Berth allocation and docking control are important aspects for the Coast Guard surveillance activities. |
| A.4<br>Suspicious Vessel / Maneuver | Accurate Vessel Positioning (terrestrial and satellite), Vessel-Port bi-directional communication | Data aggregation and on-line analytical processing | COAST GUARD | Satellite information can help preventing suspicious maneuvering and route changes |
| | | | INSURANCE COMPANIES | Adopting smart systems to prevent suspicious vessel maneuvers, in order to calculate an adequate insurance premium and to prevent extra-costs deriving from insurance indemnities |
| B.2<br>Gate Automation | Accounting for users, vehicles and goods | Integration with Gate Transit System | HAULIERS | Gate automation based on an authorization system |
| B.4<br>Freight Routing | Port-to-Port, Port-to-Road, Port-to-Railways communications | Moving from POC to full-scale deployment | INSURANCE COMPANIES | Port-to-port, port-to-road and port-to-railways communications can prevent insurance claims connected to a wrong freight routing plan |
| | | | PORT AUTHORITY | Real time information and statistics about the transport network status are important for the port activities planning operations |
| | | | CUSTOM OFFICES | An accurate freight routing is useful to reconstruct the path followed by all the specific cargo shipping lots along the supply chain |
| B.5<br>Incident at Landside | Distributed monitoring network | Data aggregation, and on-line analytical processing | HAULIERS | Accident prevention and optimized vehicle routing |
| | | | INSURANCE COMPANIES | A distributed monitoring network, connecting the ocean domain to the landside by the IoT paradigm, is used to mitigate several logistics insurance risks |
| C.1<br>Infomobility and journey monitor | Journey planner and manager (booking, payment), JIT information delivery | Missing MaaS platform | TOURISM OPERATORS | Info-mobility real-time information, journey planning and trip management are value-added service which can be offered to a tourist |
| | | | TOURISTS | MaaS and info-mobility travel planner, integrated tourism platform, including both transport information, tourism points of interest and accomodations |
| | | | COMMUTERS | MaaS and info-mobility travel planner, with a full network information (traffic jams, line interruption and maintenance etc.) is a valid ICT tool for all the people who need to reach the workplace everyday, using both private and public transport |
| | | | PUBLIC TRANSPORT | Info-mobility platforms and MaaS apps allow data aggregation and offer tailor-made services to all the passengers (both commuters and tourists). A full integration with TCC is strongly recommended |
| C.2<br>Integration with TCC | Port-to-road full-fledged data exchange | Missing a standard adaptation protocol from DATEX to C-ITS | TOURISM OPERATORS | Info-mobility real-time information, journey planning and trip management are value-added service which can be offered to a tourist |
| | | | TOURISTS | The integration with TCC is useful to provide real time information |
| | | | INSURANCE COMPANIES | A distributed monitoring network , connecting the ocean domain to the landside by the IoT paradigm is used to mitigate several logistics insurance risks. In this sense, an integration |

| | | | | with the Traffic Control Centre is required. |
|---|---|---|---|---|
| | | | PUBLIC TRANSPORT | Info-mobility platforms and MaaS apps allow data aggregation and offer tailor-made services to all the passengers (both commuters and tourists). A full integration with TCC is strongly recommended |
| C.3 In-port Smart and Autonomous Mobility (including safety) | Real-time communication Port-Vehicles-Pedestrians | Moving from POC to full-scale deployment | TOURISTS | "In-port" autonomous driving is extremely important for the pedestrian safety and for the accident prevention |
| | | | COMMUTERS | Autonomous driving, which is an important safety and security aspect in port areas, helps to prevent accidents and traffic jams |
| | | | INSURANCE COMPANIES | Seaport areas represent dangerous places for pedestrians and drivers. In this sense, autonomous driving and in-port smart mobility can be a valid instrument for mitigating several risk factors linked to safety and security in port areas |
| D.1 Pollution Level (including COx and noise) | Distributed monitoring network | Data aggregation, and on-line analytical processing | CITIZENS | Pollution level and noise reduction contribute to increase the citizen well-being. |
| | | | PORT AUTHORITY | Using pollution control systems and dynamic service pricing based on how green the processes are, environmental goals are easier to achieve |
| D.2<br>Road Traffic Level | Distributed monitoring network | Data aggregation, and on-line analytical processing | PORT AUTHORITY | Using pollution control systems and dynamic service pricing based on how green the processes are, environmental goals are easier to achieve |
| | | | CITIZENS | Reduced road traffic (due to a heavy vehicles traffic reduction) |
| **DAY 2** | | | | |
| A.2<br>Vessel maneuvering in port waters | Accurate Vessel Positioning (terrestrial and satellite), Accurate Bathymetric Data, Real-Time meteo-marine monitoring, HD video sources on vessel & port. | High-Rate/Real-Time Vessel-Port bi-directional communication | INSURANCE COMPANIES | A smart ship (real-time) control platform will mitigate the risk of failures on the ship and caused by the ship |
| | | | SHIPPING COMPANIES | Vessel maneuvering in port data is based on several aggregated and integrated data (such as meteo-marine, bathymetric, satellite and HD video sources) |
| | | | COAST GUARD | Autonomous ships and autonomous maneuverings in port water are important aspects according to the coast monitoring and control activities. This kind of "smart-navigation" better meets safety and security navigation requirements |
| D.3<br>Dynamic pricing (all services) to Vessels, Terminals | Distributed monitoring network | Data aggregation, and on-line analytical processing, data mining and knowledge extraction | CITIZENS | According to special tariffs based on the green impact of the logistics chain operators, citizens can have indirect benefits in terms of air and water pollution reduction |
| | | | HAULIERS | Hauliers can benefit of a dynamic pricing system and obtain an ad-hoc service tariff based on the environmental impact |
| | | | PORT AUTHORITY | Dynamic "ecological tariffs" are fundamental factors to improve the port green impact |

27R<small>EFERENCES</small>

[1] "J-Lab Ports website" [Online]. Available: https://jlab-ports.cnit.it/ Accessed on: Sep. 13, 2021.
[2] ISO Technical Committee (TC) 204 Intelligent Transport Systems (ITS) [Online]. Available: https://www.iso.org/committee/54706.html. Accessed on: Sep. 13, 2021.
[3] ETSI Technical Committee (TC) Intelligent Transport Systems (ITS) [Online]. Available: https://www.etsi.org/committee/its. Accessed on: Sep. 13, 2021.
[4] CEN Technical Committee (TC) Intelligent Transport Systems (ITS) [Online]. Available: https://www.itsstandards.eu/. Accessed on: Sep. 13, 2021.
[5] ETSI Technical Committee (TC) Intelligent Transport Systems (ITS), EN 302 665 "Communications Architecture", 2010.
[6] ISO/TC 204 Intelligent transport systems, ISO IS 21217 "Station and communication architecture", 2020.
[7] H.J. Fischer, "Cooperative ITS: The SDO Perspective for Early Deployment" in "Intelligent Transportation Systems - From Good Practices to Standards", P. Pagano Ed., pages 131-150. Taylor & Francis Group, Boca Raton, FL (USA): CRC Press, ISBN: 9781498721868. eBook ISBN: 978-1-4987-2187-5. DOI: 10.1201/9781315370866
[8] The Connecting Europe Facility Programme [Online]. Available: https://ec.europa.eu/inea/en/connecting-europe-facility.
[9] TENtec Interactive Map Viewer [Online]. Available: https://ec.europa.eu/transport/infrastructure/tentec/tentec-portal/map/maps.html
[10] L. Heilig and S. Voß, "Information systems in seaports: a categorization and overview," *Information Technology and Management, Springer*, vol. 18(3), pages 179-201, September 2017.
[11] UNECE Trade Facilitation and E-business (UN/CEFACT) [Online]. Available: https://unece.org/trade/uncefact. Accessed on: Sep. 13, 2021.
[12] ISO Technical Committee (TC) 8 Ships and marine technology [Online]. Available: https://www.iso.org/committee/45776.html. Accessed on: Sep. 13, 2021.
[13] ISO Technical Committee (TC) 104 Freight Containers [Online]. Available: https://www.iso.org/committee/51156.html. Accessed on: Sep. 13, 2021.
[14] IMO International Convention for the Safety of Life at Sea (SOLAS) [Online]. Available: https://www.imo.org/en/KnowledgeCentre/ConferencesMeetings/Pages/SOLAS.aspx. Accessed on: Sep. 13, 2021.
[15] Digital Container Shipping Association (DCSA) [Online]. Available: https://dcsa.org/. Accessed on: Sep. 13, 2021.
[16] ETSI Radio Spectrum Matters (RSM) Task Group (TG) Marine [Online]. Available: https://portal.etsi.org/TB-SiteMap/ERM/ERM-ToR/ERMtg26-ToR. Accessed on: Sep. 13, 2021.
[17] ETSI ISG on European Common Information Sharing Environment Service and Data Model (CDM) [Online]. Available: https://www.etsi.org/committee/1584-cdm. Accessed on: Sep. 13, 2021.
[18] P. Krugman, "Increasing Returns and Economic Geography", 1991, Journal of Political Economy [Online] Available: https://pr.princeton.edu/pictures/g-k/krugman/krugman-increasing_returns_1991.pdf. Accessed on Sep. 5, 2021.
[19] A. Molavi, G. J. Lim, B. Race "A Framework for Building a Smart Port and Smart Port Index" [Online] Available: https://www.researchgate.net/publication/332684618_A_Framework_for_Building_a_Smart_Port_and_Smart_Port_Index.
[20] A. R. González, N. González-Cancelas, B. Molina Serrano, A. Camarero "Preparation of a Smart Port Indicator and Calculation of a Ranking for the Spanish Port System" [Online] Available: https://www.researchgate.net/publication/341126423_Preparation_of_a_Smart_Port_Indicator_and_Calculation_of_a_Ranking_for_the_Spanish_Port_System. Accessed on: Jul. 26, 2021.
[21] "Directive 2002/59 of the European Parliament"
[22] "Regulation 2016/399 of the European Parliament"
[23] "Directive 2002/59 of the European Parliament"
[24] "Directive 2019/883 of the European Parliament"
[25] "Regulation 2004/725 of the European Parliament"
[26] "Regulation 2013/952 of the European Parliament"
[27] A. Moros-Daza, R. Amaya-Mier and C. Paternina-Arboleda, "Port Community Systems: A structured literature review," *Transportation Research Part A: Policy and Practice, Elsevier*, vol. 133, pages 27-46, 2020. ISSN 0965-8564.
[28] Network of Trusted Networks (NoTN) [Online]. Available: https://ipcsa.international/initiatives/network-of-trusted-networks/. Accessed on: Sep. 13, 2021.
[29] "TradeLens | Digitizing the global supply chain and transforming trade." [Online]. Available: https://www.tradelens.com. Accessed on: Sep. 13, 2021.
[30] A. Guidi "The watchwords: technology and smart ports" [Online]. Available: https://www.ispionline.it/it/pubblicazione/watchwords-technology-and-smart-ports-29823. Accessed on Aug. 24, 2021.
[31] R. Philipp "Digital readiness index assessment towards smart port development" [Online] Available: https://link.springer.com/article/10.1007%2Fs00550-020-00501-5. Accessed on Jul. 26, 2021.
[32] Banca d'Italia "Le infrastrutture in Italia: dotazione, programmazione, realizzazione" [Online] Avaliable: https://www.bancaditalia.it/pubblicazioni/collana-seminari-convegni/2011-0007/7_infrastrutture_italia.pdf. Accessed on Jul. 26, 2021.
[33] T. Jouili "The Role of Seaports in the Process of Economic Growth" [Online] Available: https://core.ac.uk/download/pdf/234682718.pdf. Accessed on Jul. 26, 2021.
[34] W. T. Talley, "Port economics", ", London: Routledge, 2009 [Online]. Available: https://www.taylorfrancis.com/books/mono/10.4324/9781315667720/port-economics-talley-wayne. Accessed on Jul. 26, 2021.
[35] M. Stopford, "Maritime Economy", London: Routledge, 2009 [Online] Available: https://www.taylorfrancis.com/books/mono/10.4324/9780203891742/maritime-economics-3e-martin-stopford. Accessed on Aug. 5, 2021.
[36] P. L. Pallis, "Port Risk Management in Container Terminals", 2017, Transportation Research Procedia [Online] Available: https://www.researchgate.net/publication/317418602_Port_Risk_Management_in_Container_Terminals. Accessed on Aug 10, 2021.
[37] TM2.0, "Mobility as a Service Task Force – Final Report," June 2019 [Online]. Available: http://tm20.org/wp-content/uploads/sites/8/2019/08/TM2.0-TF_MaaS_Final_Report_v3.0.pdf/. Accessed on: Sep. 13, 2021.